\documentclass[prl,aps,showpacs,epsf, twocolumn]{revtex4}

\usepackage{graphicx}
\usepackage{epsfig}

\begin{document}

\title{Experimental Study of the BEC-BCS Crossover Region in Lithium 6}
\author{T.\,Bourdel, L.\,Khaykovich, J.\,Cubizolles, J.\,Zhang, F.\,Chevy,
M.\,Teichmann, L.\,Tarruell, S.\,J.\,J.\,M.\,F.\,Kokkelmans, and
C.\,Salomon} \affiliation{ Laboratoire Kastler-Brossel, ENS, 24
rue Lhomond, 75005 Paris} \date{\today}

\begin{abstract}
We report Bose-Einstein condensation of weakly bound $^6$Li$_2$
molecules in a crossed optical trap near a Feshbach resonance. We
measure a molecule-molecule scattering length of
$170^{+100}_{-60}$\,nm at 770\,G, in good agreement with theory.
We study the 2D expansion of the cloud and show deviation from
hydrodynamic behavior in the BEC-BCS crossover region.
\end{abstract}


\maketitle

By applying a magnetic field to a gas of ultra-cold atoms, it is
possible to tune the strength and the sign of the effective
interaction between particles. This phenomenon, known as Feshbach
resonance, offers in the case of fermions the unique possibility
to study the crossover between situations governed by
Bose-Einstein and Fermi-Dirac statistics. Indeed, when the
scattering length $a$ characterizing the 2-body interaction at low
temperature is positive, the atoms are known to pair in a bound
molecular state. When the temperature is low enough, these bosonic
dimers can form a Bose-Einstein condensate (BEC) as observed very
recently in $^{40}$K \cite{Greiner03} and $^6$Li
\cite{Jochim03,Zwierlein03}. On the side of the resonance where
$a$ is negative, one expects the well known
Bardeen-Cooper-Schrieffer (BCS) model for superconductivity to be
valid. However, this  simple picture of a BEC phase on one side of
the resonance and a BCS phase on the other is valid only for small
atom density $n$. When $n|a|^3\gtrsim 1$ the system enters a
strongly interacting regime that represents a challenge for
many-body theories \cite{Leggett80,Heiselberg01,Carlson03}
and that now begins to be accessible to experiments
\cite{Bartenstein04,Regal04,Zwierlein04}.

In this letter, we report on Bose-Einstein condensation of $^6$Li
dimers in a crossed optical dipole trap and a study of the BEC-BCS
crossover region. Unlike all previous observations of molecular
BEC made in single beam dipole traps with very elongated
geometries, our condensates are formed in nearly isotropic
 traps. Analyzing
 free expansions of pure condensates with up to $4\times 10^4$ molecules,
 we measure the molecule-molecule scattering length
$a_{\rm m}=170^{+100}_{-60}$\,nm at a magnetic field of 770\,
gauss. This measurement is in good agreement with the value
deduced from the resonance position \cite{Zwierlein04} and the
relation $a_{\rm m}=0.6\,a$\,of ref.\,\cite{Petrov03}. Combined
with tight confinement, these large scattering lengths lead to a
regime of strong interactions where the chemical potential $\mu$
is on the order of $k_{\rm B} T_{\rm C}$ where  $T_{\rm C}\simeq
1.5\,\mu$K is the condensation temperature. As a consequence, we
find an important modification of the thermal cloud time of flight
expansion induced by the large condensate mean field. Moreover,
the gas parameter $n_{\rm m}a_{\rm m}^3$ is no longer small but on
the order of $0.3$. In this regime, the validity of mean field
theory becomes questionable \cite{Baym01,Dalfovo99,Gerbier04}. We
show, in particular, that the anisotropy and gas energy released
during expansion varies monotonically across the Feshbach
resonance.

Our experimental setup has been described previously
\cite{Bourdel03, Cubizolles03}. A gas of $^6$Li atoms is prepared
in the absolute ground state $|1/2,1/2\rangle$ in a Nd-YAG crossed
beam optical dipole trap. The horizontal beam (resp.\,vertical)
propagates along $x$ ($y$), has a maximum power of $P_o^h=2\,$W
($P_o^v=3.3\,$W) and a waist of $\sim 25\,\mu$m ($\sim 40\,\mu$m).
At full power, the $^6$Li trap oscillation frequencies are
$\omega_x/2\pi= 2.4(2)\,$kHz, $\omega_y/2\pi=5.0(3)\,$kHz, and
$\omega_z/2\pi=5.5(4)$\,kHz, as measured by parametric excitation,
and the trap depth is $\sim 80\,\mu$K. After sweeping the magnetic
field $B$ from 5\,G to 1060\,G, we drive the Zeeman transition
between $|1/2,1/2\rangle$ and $|1/2,-1/2\rangle$ with a 76\,MHz RF
field to prepare a balanced mixture of the two states. As measured
very recently \cite{Zwierlein04}, the Feshbach resonance between
these two states is peaked at 822(3)\,G, and for $B$=1060\,G,
$a=-167$\,nm. After 100\,ms the coherence between the two states
is lost and plain evaporation provides
$N_\uparrow=N_\downarrow=N_{\rm tot}/2=1.5\times 10^5$ atoms at
10\,$\mu$K=0.8\,$T_{\rm F}$, where $k_{\rm B}T_{\rm
F}=\hbar^2k_{\rm F}^2/2m=\hbar(3 N_{\rm tot}\omega_x \omega_y
\omega_z)^{1/3}=\hbar\bar{\omega}(3 N_{\rm tot})^{1/3}$ is the
Fermi energy. Lowering the intensity of the trapping laser to
$0.1\,P_0$, the Fermi gas is evaporatively cooled to temperatures
$T$ at or below $0.2\,T_{\rm F }$ and $N_{\rm tot}\approx 7\times
10^4$.

\begin{figure}[hb]
\begin{center}
\epsfig{file=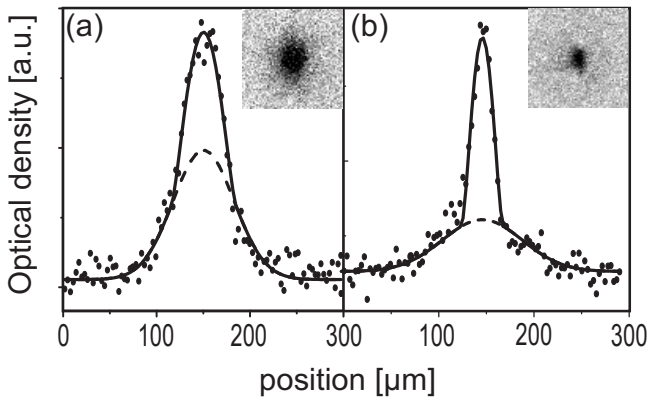, width=\linewidth}
\caption{\label{fig:figure1} Onset of Bose-Einstein condensation
in a cloud of $2\times 10^4$ $^6$Li dimers at 770\,G (a) and of
$2\times 10^4$ $^7$Li atoms at 610\,G (b) in the same optical
trap. (a): 1.2\,ms expansion profiles along the weak direction $x$
of confinement. (b): 1.4 ms expansion. The different sizes of the
condensates reflect the large difference in scattering length
$a_{\rm m}=170\,$nm for $^6$Li dimers and $a_{7}=0.55$\,nm for
$^7$Li. Solid line: Gaussian+Thomas-Fermi fit. Dashed line:
gaussian component. Condensate fractions are both $28\,\%$.
$\omega_x/2\pi= 0.59(4)\,$kHz, $\omega_y/2\pi=1.6(1)\,$kHz, and
$\omega_z/2\pi=1.7(1)$\,kHz in (a). $\omega_x/2\pi= 0.55(4)\,$kHz,
$\omega_y/2\pi=1.5(1)\,$kHz, and $\omega_z/2\pi=1.6(1)$\,kHz in
(b). }
\end{center}
\end{figure}
Then, sweeping  the magnetic field to $770\,$G in $200\,$ms, the
Feshbach resonance is slowly crossed. In this process atoms are
reversibly transformed into cold molecules
\cite{Cubizolles03,Regal03} near the BEC critical temperature as
presented in figure 1a.  The onset of condensation is revealed by
bimodal and anisotropic momentum distributions in time of flight
expansions of the molecular gas. These images are recorded as
follows. At a fixed magnetic field, the optical trap is first
switched off. The cloud expands typically for 1\,ms and then the
magnetic field is increased by 100 \,G in $50\,\mu$s.This converts
the molecules back into free atoms above resonance without
releasing their binding energy \cite{Zwierlein03}. Switching the
field abruptly off in $10\,\mu$s, we detect free $^6$Li atoms by
light absorption near the D2 line. Using this method, expansion
images are not altered by the adiabatic following of the molecular
state to a deeper bound state during switch-off as observed in our
previous work \cite{Cubizolles03}. Furthermore, we check that
there are no unpaired atoms before expansion. In figure 1b, a
Bose-Einstein condensate of $^7$Li atoms produced in the same
optical trap is presented. The comparison between the condensate
sizes after expansion
reveals that the
mean field interaction and scattering length are much larger for
$^6$Li$_2$ dimers (Fig.\,1a) than for $^7$Li atoms (Fig.\,1b).

To measure the molecule-molecule scattering length, we produce
pure molecular condensates by taking advantage of our crossed
dipole trap. We recompress the horizontal beam to full power while
keeping the vertical beam at the low power of 0.035\,$P_0^v$
corresponding to a trap depth for molecules $U=5.6\,\mu$K.
Temperature is then limited to $T\leq 0.9\,\mu$K assuming a
conservative $\eta=U/k_{\rm B}T=6$, whereas the critical
temperature increases with the mean oscillation frequency.
Consequently, with an axial (resp. radial) trap frequency of
440\,Hz (resp. 5\,kHz), we obtain $T/T_{\rm C}^0\leq 0.3$, where
$T_{\rm C}^0=\hbar\bar{\omega}(0.82N_{\rm
tot}/2)^{1/3}=$2.7\,$\mu$K is the non interacting BEC critical
temperature. Thus, the condensate should be pure as confirmed by
our images.
 After 1.2~ms of expansion, the radius of the
condensate in the $x$  (resp. $y$) direction is $R_x=51~\mu$m
($R_y=103~\mu$m). The resulting anisotropy $R_y/R_x=2.0(1)$ is
consistent with the value 1.98 \cite{correctioncourbure} predicted
the scaling equations \cite{Kagan96,Castin96}. Moreover, this set
of equation  leads to an {\em in-trap} radius $R_x^0=26\mu$m
(resp. $R_y^0=2.75\mu$m). We then deduce the molecule-molecule
scattering length from the Thomas-Fermi formula $R_{x,y}^0= a_{\rm
ho}\bar\omega/\omega_{x,y}(15 N_{\rm tot}a_{\rm m}/2a_{\rm
ho})^{1/5}$, with $a_{\rm ho}=\sqrt{\hbar/2m\bar\omega}$.
Averaging over several images, this yields $a_{\rm
m}=$170$^{+100}_{-60}$\,nm at 770\,G,. Here, the statistical
uncertainty is negligible compared to  the systematic uncertainty
due to the calibration of our atom number.  At this field, we
calculate an atomic scattering length of $a= 306$\,nm. Combined
with the prediction $a_{\rm m}=0.6\,a$ of \cite{Petrov03},  we
obtain $a_{\rm m}=183\,$nm in good agreement with our measurement.
For $^7$Li, we obtain with the same analysis a much smaller
scattering length of $a_7$=0.65(10)\,nm at 610\,G also in
agreement with theory \cite{Khaykovich02}.

Such large values of $a_{\rm m}$ bring our molecular condensates
into a novel regime where the gas parameter $n_{\rm m}a_{\rm m}^3$
is no longer very small. Indeed, $a_{\rm m}= 170$\,nm and
$n_{\rm m}=6\times 10^{13}$cm$^{-3}$ yield $n_{\rm m}a_{\rm
m}^3=0.3$. As a first consequence, corrections due to beyond mean
field effects \cite{Pitaevskii98,Baym01} or to the underlying
fermionic nature of atoms may play a role, since the average
spacing between molecules is then of the order of the molecule
size $\sim a/2$. Second, even in a mean field approach,
thermodynamics is expected to be modified. For instance, in the
conditions of Fig.\,1a, we expect a large shift of the BEC
critical temperature \cite{Baym01,Dalfovo99,Gerbier04}. The shift
calculated to first order in $n^{1/3}a$ \cite{Dalfovo99}, $\Delta
T_{\rm C}/T_{\rm C}^0=-1.4$,  is clearly inapplicable and a more
refined approach is required \cite{Kokkelmans04}. Third, we
observe that partially condensed cloud expansions are modified by
interactions. Indeed, double structure fits lead to temperatures
inconsistent with the presence of a condensate. In Fig.\,1, we
find $T=1.6\,\mu$K, to be compared with $T_{\rm C}^0=1.4\,\mu$K,
whereas for the $^7$Li condensate $T=0.7\,\mu$K=0.6$T_{\rm C}^0$.

This inconsistency results from the large mean field interaction
which modifies the thermal cloud expansion. To get a better
estimate of the temperature, we rely on a release energy
calculation. We calculate the Bose distribution of thermal atoms
in a mexican hat potential that is the sum of the external
potential and the repulsive mean field potential created by the
condensate. For simplicity we neglect the mean field resulting
from the thermal component. The release energy is the sum of the
thermal kinetic energy, condensate interaction energy, and
Hartree-Fock interaction energy between the condensate and thermal
cloud. The temperature and chemical potential are then adjusted to
fit the measured atom number and release energy. For figure 1a, we
obtain a condensate fraction of $28\,\%$ and $\mu=\hbar
\bar{\omega}/2(15N_{tot}a_{\rm m}/2a_{\rm ho})^{2/5}=0.45\,\mu$K.
The temperature $T=1.1\,\mu$K is then found below $T_{\rm
C}^0=1.4\,\mu$K.

The condensate lifetime is typically $\sim$300\,ms at 715\,G
($a_{\rm m}=66\,$nm) and $\sim$3\,s at 770\,G ($a_{\rm
m}=170\,$nm), whereas for $a=-167$\, nm at 1060\,G, the lifetime
exceeds 30\,s. On the BEC side, the molecule-molecule loss rate
constant is $G=0.26^{+0.08}_{-0.06} \times 10^{-13}\,$cm$^3$/s at
770\,G and $G=1.75^{+0.5}_{-0.4} \times 10^{-13}\,$cm$^3$/s at
715\,G with the fit procedure for condensates described in
\cite{Soding97}.  Combining similar results for four values of the
magnetic field ranging from 700~G to 770~G, we find $G\propto
a^{-1.9 \pm 0.8}$. Our data are in agreement with the theoretical
prediction $G \propto a^{-2.55}$ of ref.\,\cite{Petrov03} and with
 previous measurements of G in a thermal gas at 690\,G
\cite{Cubizolles03} or in a BEC at 764~G \cite{Bartenstein04}. A
similar power law was also found for $^{40}$K \cite{Regal04b}.

We now present an investigation of the crossover from a Bose-Einstein
condensate to an interacting Fermi gas (Fig.\,2 and 3). We prepare a
nearly pure condensate with 3.5$\times 10^4$ molecules at 770\,G and
recompress the trap to frequencies of $\omega_x=2 \pi \times 830\,$Hz,
$\omega_y=2 \pi \times 2.4\,$kHz, and $\omega_z=2 \pi \times
2.5\,$kHz. The magnetic field is then slowly swept at a rate of
2\,G/ms to various values across the Feshbach resonance. The 2D
momentum distribution after a time of flight expansion of 1.4\,ms is
then detected as previously.

\begin{figure}[ht]
\begin{center}
\epsfig{file=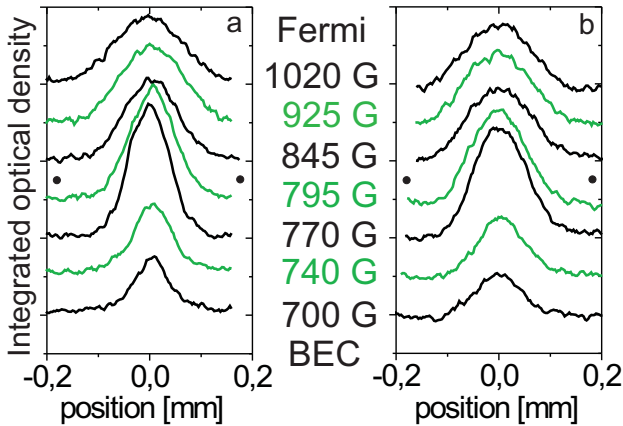, width=0.99\linewidth} \caption{
\label{fig:figure4} Integrated density profiles across the BEC-BCS
crossover region. 1.4\,ms time of flight expansion in the axial
(a) and radial (b) direction. The magnetic field is varied over
the whole region of the Feshbach resonance from $a>0$ ($B<810\,$G)
to $a<0$ ($B>810\,$G). $\bullet$:\,Feshbach resonance peak.}
\end{center}
\end{figure}
Fig.\,2 presents the observed profiles (integrated over the
orthogonal direction) for different values of the magnetic field.
At the lowest field values $B\leq 750\,$G, $n_{\rm m}a_{\rm m}^3
\ll 1$, condensates number are relatively low because of the
limited molecule lifetime. As $B$ increases, the condensate width
gradually increases towards the width of a non interacting Fermi
gas, and nothing dramatic happens on resonance. At the highest
fields (B$\geq 925$\,G), where $k_{\rm F} |a| \leq 3$,
distributions are best fitted with zero temperature Fermi
profiles. More quantitatively, Fig.\,3b presents both the gas
energy released after expansion $E_{\rm rel}$ and the anisotropy
$\eta$ across resonance.  These are calculated from gaussian fits
to the density after time of flight: $E_{\rm rel}=m
(2\sigma_y^2+\sigma_x^2)/2\tau^2$ and $\eta=\sigma_y/\sigma_x$,
where $\sigma_i$ is the rms width along $i$, and $\tau$ is the
time of flight \cite{correctioncourbure}. On the BEC side at 730\,
G, the measured anisotropy is $\eta \sim$1.6(1) , in agreement
with the hydrodynamic  prediction, 1.75. It then decreases
monotonically to 1.1 at 1060\,G on the BCS side. On resonance, at
zero temperature, superfluid hydrodynamic expansion is expected
\cite{Menotti02} corresponding to  $\eta=1.7$.  We find however
$\eta=1.35(5)$, indicating a partially hydrodynamic behavior that
could be due to a reduced superfluid fraction. On the $a<0$ side,
the decreasing anisotropy would indicate a further decrease of the
superfluid fraction that could correspond to the reduction of the
condensed fraction of fermionic atom pairs away from resonance
observed in \cite{Regal04,Zwierlein04}. Interestingly, our results
differ from that of ref.\cite{OHara02} where hydrodynamic
expansion was observed at $910\,$G in a more elongated trap for
$T/T_{\rm F}\simeq 0.1$.


\begin{figure}[htb]
\begin{center}
\epsfig{file=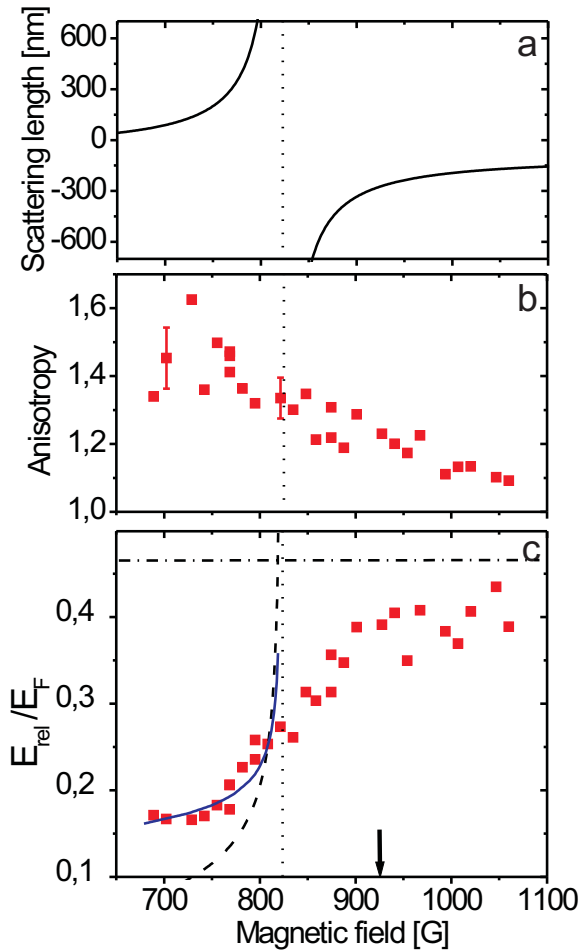, width=0.99\linewidth} \caption{
\label{fig:figure3} (a): scattering length between the
$|1/2,1/2\rangle$ and $|1/2,-1/2\rangle$ $^6$Li states. The
Feshbach resonance peak is located at 820\,G (dotted line). (b):
anisotropy of the cloud,(c): release energy across the BEC-BCS
crossover region. In (c), the dot-dashed line corresponds to a
$T=0$ ideal Fermi gas. The dashed curve is the release energy from
a pure condensate in the Thomas-Fermi limit. The solid curve
corresponds to a finite temperature mean field model described in
the text with $T=0.5\,T_{\rm C}^0$. Arrow: $k_{\rm F}|a|=3$.}
\end{center}
\end{figure}

In the BEC-BCS crossover regime, the gas energy released after
expansion $E_{\rm rel}$ is also smooth (Fig.\,3c). $E_{\rm rel}$
presents a plateau for $B\leq 750\,$G, and then increases
monotonically towards that of a weakly interacting Fermi gas. The
plateau is not reproduced by the mean field approach of a pure
condensate (dashed line). This is a signature that the gas is not
at $T=0$. It can be understood with the mean field approach we
used previously to describe the behavior of the thermal cloud.
Since the magnetic field sweep is slow compared to the gas
collision rate \cite{Cubizolles03}, we assume that this sweep is
adiabatic and conserves entropy \cite{Carr03}. We then adjust this
entropy to reproduce the release energy at a particular magnetic
field, $B=720$\,G. The resulting curve as a function of $B$ (solid
line in Fig.\,3c) agrees well with our data in the range
$680\,$G$\,\leq B\leq 800\,$G, where the condensate fraction is
70$\%$, and the temperature is $T\approx T_{\rm
C}^0/2=\,1.2\,\mu$K. This model is limited to $n_ma_{\rm m}^3
\lesssim 1$. Near resonance the calculated release energy diverges
and clearly departs from the data. On the BCS side, the release
energy of a $T=0$ ideal Fermi gas gives an upper bound for the
data (dot-dashed curve), as expected from negative interaction
energy and a very cold sample. This low temperature is supported
by our measurements on the BEC side and the assumption of entropy
conservation through resonance which predicts $T=0.1\,T_{\rm F}$
\cite{Carr03}.

 On resonance the gas is expected to reach a universal
behavior, as the scattering length $a$ is not a relevant parameter
any more \cite{Heiselberg01}. In this regime, the release energy
scales as $E_{\rm rel}=\sqrt{1+\beta} E_{\rm rel}^0$, where
$E_{\rm rel}^0$ is the release energy of the non-interacting gas
and $\beta$ is a universal parameter. From our data at 820~G, we
get $\beta=-0.64(15)$ . This value is larger than the Duke result
$\beta=-0.26\pm 0.07$ at 910\,G \cite{OHara02}, but agrees with
that of Innsbruck $\beta=-0.68^{+0.13}_{-0.10}$ at 850\,G
\cite{Bartenstein04}, and with the most recent theoretical
prediction $\beta=-0.56$ \cite{Carlson03}.
Around 925\,G, where $a=-270\,$nm and $(k_{\rm F}|a|)^{-1}=0.35$,
the release energy curve displays a change of slope. This is a
signature of the transition between the strongly and weakly
interacting regimes. It is also observed near the same field in
\cite {Bartenstein04} through {\it in situ} measurement of the
trapped cloud size. Interestingly, the onset of resonance
condensation of fermionic atom pairs observed in $^{40}$K
\cite{Regal04} and $^{6}$Li \cite{Zwierlein04}, corresponds to a
similar value of $k_{\rm F}|a|$.

In summary, we have explored the whole region of the $^6$Li
Feshbach resonance, from a Bose-Einstein condensate of fermion
dimers to an ultra-cold interacting Fermi gas. The extremely large
scattering length between molecules, that we have measured leads
to novel BEC conditions. We have observed hydrodynamic expansions
on the BEC side and non-hydrodynamic expansions at and above
resonance. We suggest that this effect  results from a reduction
of the superfluid fraction and we point to the need of a better
understanding of the dynamics of an expanding Fermi gas.


We are grateful to Y.\,Castin, C.\, Cohen-Tannoudji,
R.\,Combescot, J.\,Dalibard, G.\,Shlyapnikov, and S.\,Stringari
for fruitful discussions. This work was supported by CNRS,
Coll\`ege de France, and R\'egion Ile de France. S.\,Kokkelmans
acknowledges a Marie Curie grant from the E.U. under contract
number MCFI-2002-00968. Laboratoire Kastler Brossel is {\it
Unit\'e de recherche de l'Ecole Normale Sup\'erieure et de
l'Universit\'e Pierre et Marie Curie, associ\'ee au CNRS}.


\begin{thebibliography}{99}
\bibitem{Greiner03}
M.\,Greiner, C.\,A.\,Regal, and D.\,S\,Jin, Nature {\bf 426}, 537 (2003).
\bibitem{Jochim03}
S.\,Jochim {\it et al.}, Science {\bf 302}, 2101 (2003).
\bibitem{Zwierlein03}
M.\,W.\,Zwierlein {\it et al.}, Phys. Rev. Lett. {\bf 91}, 250401
(2003).
\bibitem{Leggett80}
A.J. Leggett, J. Phys. C. (Paris) {\bf 41},7 (1980);  P.
Nozi\`eres and S. Schmitt-Rink, J. Low Temp. Phys. {\bf 59} 195
(1985); C. S\'a de Melo, M. Randeria, and J. Engelbrecht, Phys.
Rev. Lett. {\bf 71}, 3202 (1993);  M.\,Holland, S.\,Kokkelmans,
M.\,Chiofalo, and R.\,Walser, Phys. Rev. Lett. {\bf 87} 120406
(2001); Y.\,Ohashi and A.\,Griffin, Phys. Rev. Lett.  {\bf 89},
130402 (2002); J.\,N.\,Milstein, S.\,J.\,J.\,M.\,F.\,Kokkelmans,
and M.\,J.\,Holland, Phys. Rev. A {\bf 66}, 043604 (2002); R.
Combescot, Phys. Rev. Lett. {\bf 91}, 120401 (2003); G.\,M.\,Falco
and H.\,T.\,C.\,Stoof, cond-mat/0402579.
\bibitem{Heiselberg01} H. Heiselberg, Phys. Rev. A {\bf 63}, 043606
(2001).
\bibitem{Carlson03}
J.\,Carlson, S-Y\,Chang, V.\,R.\,Pandharipande, and
K.\,E.\,Schmidt, Phys. Rev. Lett. {\bf 91}, 050401 (2003).
\bibitem{Regal04}
C.\,A\, Regal, M.\,Greiner, and D.\,S.\,Jin
Phys. Rev. Lett. {\bf 92}, 040403 (2004).
\bibitem{Bartenstein04}
M.\,Bartenstein {\it et al.}, cond-mat/0401109.
\bibitem{Zwierlein04}
M. Zwierlein {\it et al.}, cond-mat/0403049.
\bibitem{Petrov03} D.\,S.\,Petrov, C.\,Salomon, and G.\,V.\,Shlyapnikov, cond-mat/0309010
(2003).

\bibitem{Baym01} G. Baym {\it et al.}, EPJ B {\bf 24}, 107 (2001)
\bibitem{Dalfovo99}
F.\,Dalfovo, S.\,Giorgini, L.\,P.\,Pitaevskii,
and S.\,Stringari, Rev. of Mod. Phys.  {\bf 91}, 463 (1999).
\bibitem{Gerbier04}
F. Gerbier {\it et al.}, Phys. Rev. Lett. {\bf 92}, 030405 (2004).
\bibitem{Cubizolles03} J.\,Cubizolles {\it et al}, Phys. Rev. Lett. {\bf 91} 240401 (2003).
\bibitem{Bourdel03}
T.\,Bourdel {\it et al.}, Phys. Rev. Lett. {\bf 91}, 020402
(2003).
\bibitem{Regal03}
C.\,A.\,Regal, C.\,Ticknor, J.\,L.\,Bohn, and D.\,S.\,Jin, Nature {\bf 424}, 47 (2003).
\bibitem{correctioncourbure} We correct our data for the presence of
a magnetic field curvature which leads to an anti-trapping
frequency of 100\,Hz at 800\,G along $x$.
\bibitem{Kagan96}
Yu.\,Kagan, E.\,L.\,Surkov, and G.\,V.\,Shlyapnikov, Phys. Rev. A
{\bf 54}, R1753 (1996).
\bibitem{Castin96}
Y.\,Castin and R.\,Dum, Phys. Rev. Lett. {\bf 77}, 5315 (1996).
\bibitem{Khaykovich02}
 L.\,Khaykovich {\it et al.}, Science {\bf 296}, 1290 (2002).
\bibitem{Pitaevskii98}
L. Pitaevskii and S. Stringari, Phys. Rev. Lett. {\bf 81},
4541(1998).
\bibitem{Kokkelmans04}A mean field self-consistent calculation of the molecular density profile
in the trap at $T_{\rm C}$ leads to $T_{\rm C}^{\rm mf}=0.58T_{\rm
C}^0\approx 0.8\,\mu$K. S. Kokkelmans, to be published.
\bibitem{Soding97}
J.\,S\"oding, {\it et al}, Applied Physics {\bf B69}, 257 (1999)
\bibitem{Regal04b} C. Regal, M. Greiner and D. Jin, Phys. Rev.
Lett. {\bf 92}, 083201 (2004).
\bibitem{Menotti02}
C. Menotti, P. Pedri, and S. Stringari, Phys. Rev. Lett. {\bf 89},
250402 (2002).
\bibitem{OHara02}
K. O'Hara {\it et al.}, Science {\bf 298}, 2179 (2002). M. Gehm
{\it et al.}, Phys. Rev. A {\bf 68}, 011401 (2003).
\bibitem{Carr03}
L.\,D.\,Carr, G.\,V.\,Shlyapnikov, and Y.\,Castin,
cond-mat/0308306.
\end{thebibliography}
\end{document}